\documentclass[aps,preprint,onecolumn,showpacs,superscriptaddress,groupedaddress]{revtex4}  
\usepackage{graphicx}  
\usepackage{float}
\usepackage{dcolumn}   
\usepackage{bm}        
\usepackage{amssymb}   
\usepackage{amsmath}
\usepackage{isomath}
\usepackage[usenames, dvipsnames]{color}
\usepackage{xcolor}
\usepackage{soul}
\usepackage{physics}
\usepackage[normalem]{ulem}

\begin{document}

\title{Dial-in Topological Metamaterials \\ Based on Bistable Stewart Platform}

\author{Ying Wu}
\affiliation{Aeronautics and Astronautics, University of Washington, Seattle, WA 98195, USA}
\affiliation{Department of Astronautic Science and Mechanics, Harbin Institute of Technology, Harbin, Heilongjiang 150001, China}
\author{Rajesh Chaunsali}
\affiliation{Aeronautics and Astronautics, University of Washington, Seattle, WA 98195, USA}
\author{Hiromi Yasuda}
\affiliation{Aeronautics and Astronautics, University of Washington, Seattle, WA 98195, USA}
\author{Kaiping Yu}
\thanks{Corresponding authors: yukp@hit.edu.cn (Kaiping Yu), jkyang@aa.washington.edu (Jinkyu Yang)}
\affiliation{Department of Astronautic Science and Mechanics, Harbin Institute of Technology, Harbin, Heilongjiang 150001, China}
\author{Jinkyu Yang}
\thanks{Corresponding authors: yukp@hit.edu.cn (Kaiping Yu), jkyang@aa.washington.edu (Jinkyu Yang)}
\affiliation{Aeronautics and Astronautics, University of Washington, Seattle, WA 98195, USA}


\begin{abstract}
Recently, there have been significant efforts to guide mechanical energy in structures by relying on a novel topological framework popularized by the discovery of topological insulators. Here, we propose a topological metamaterial system based on the design of the Stewart Platform, which can not only guide mechanical waves robustly in a desired path, but also can be tuned \textit{in situ} to change this wave path at will. Without resorting to any active materials, the current system harnesses bistablilty in its unit cells, such that tuning can be performed simply by a dial-in action. Consequently, a topological transition mechanism inspired by the quantum valley Hall effect can be achieved. We show the possibility of tuning in a variety of topological and traditional waveguides in the same system, and numerically investigate key qualitative and quantitative differences between them. We observe that even though both types of waveguides can lead to significant wave transmission for a certain frequency range, topological waveguides are distinctive as they support robust, back scattering immune, one-way wave propagation.  
\end{abstract}


\maketitle

The discovery of topological insulators \citep{1,2}, which are labelled as a new state of matter in the condensed matter physics, has galvanized research efforts in multiple fields. Topological framework enables us to understand fascinating phenomena such as quantum Hall effect (QHE) \citep{3,3a}, quantum spin Hall effect (QSHE) \citep{4}, and quantum valley Hall effect (QVHE) \citep{8}. These are special because of the intriguing directional and robust edge states they support. It is recent that these fundamental concepts have been extended to the fields of photonics \citep{14}, acoustics \citep{16,17,18,19,21,22,23}, and mechanics \citep{27,28,29,30,31,32,33,34,36,38,39,40}. Such extension is of the fundamental interest to the research community as it has potential to set new design principles for \textit{topological metamaterials} that aim to strategically tailor energy transport for waveguiding, isolating, switching, filtering, and related applications.

Acoustic and mechanical metamaterials that rely on QHE need active components (e.g., gyroscope and flow circulator) or applications of external fields (e.g., magnetic field) to break the time reversal symmetry \citep{16,17,18,28,29,30,31}. This adds complexity to the system, and thus, these metamaterials based on QHE are challenging to be realized in practical environments. QSHE-inspired metamaterials employ only passive components, but usually mandate intricate ways to achieve a double Dirac cone in their dispersion relation \citep{19,21,32,33,34,36}. Metamaterials based on QVHE \citep{22,23,38,39,40}, however, rely on the breakage of the inversion symmetry to achieve topological properties, and these can be comparably easier to be designed and realized in practical settings. QVHE originates from the newly discovered valley degree of freedom (DOF) of electrons in the two-dimensional honeycomb lattice of graphene \citep{8}. These valley DOFs are energetically degenerate but are largely separated in momentum space \citep{22,23}. Due to this large separation, inter-valley scattering can be avoided, and valley DOFs constitute two pseudo-spins, which have opposite-directional and robust properties on the topological interfaces. On this principle, back-scattering immune and robust energy transport along topological waveguides with sharp bends have been proposed in acoustic \citep{22,23} and mechanical \citep{39,40} systems.

Though the aforementioned configurations have shown to be guiding energy in a specific path in the system, there remains a challenge{; Can} one change the path \textit{in situ} and thus achieve a complete control over the waveguide? Once achieved, this will provide a fertile testbed for future experiments related to wave guiding capability of various types of topological interfaces, and at the same time those can be compared with traditional waveguides in the same system more closely than ever. However, adding such versatility into the system comes at a cost. Generally, \textit{in situ} tunability requires complex components or mechanisms to be present in the system, so that the wave path in the lattice structure can be reconfigured in a controllable and versatile manner. But such complexity in design could again make the system cumbersome for practical use, and it would defeat the purpose of building a simple QVHE-based system to some extent. 

Here we show that \textit{in situ} tunability in the QVHE-based mechanical metamaterial can be achieved by utilizing nonlinearity of the constituent elements in the system. More specifically, we use an assembly of the Stewart Platform (SP), in which translation and rotational degrees of freedoms of each SP are judiciously tailored to achieve a bistable response. Consequently, a simple dial-in action changes its configuration from one stable state to the other, and this feature can be used for \textit{in situ} control of the wave path in the system. The SP already has a wide range of engineering applications such as vibration control, precise positioning, and flight simulation \citep{41,42}. Therefore, by integrating the elegant engineering of SP with the fascinating physics of QVHE, we propose a dial-in mechanical metamaterial for creating tunable topological waveguides. We use extensive numerical simulations and show that this metamaterial can be tuned \textit{in situ} to design a variety of topological waveguides with robust wave propagation characteristics. The tunability allows us to also build traditional waveguides by suppressing topological variations in the same systems, making it possible to compare their performance with the topological counterpart to a remarkable detail. Such a comparison therefore plays a key role in extending our knowledge and appreciation towards the uniqueness of topological waveguides in the proposed system. 

\section*{RESULTS}
\begin{figure}[t]
\centering
\includegraphics[width=5in]{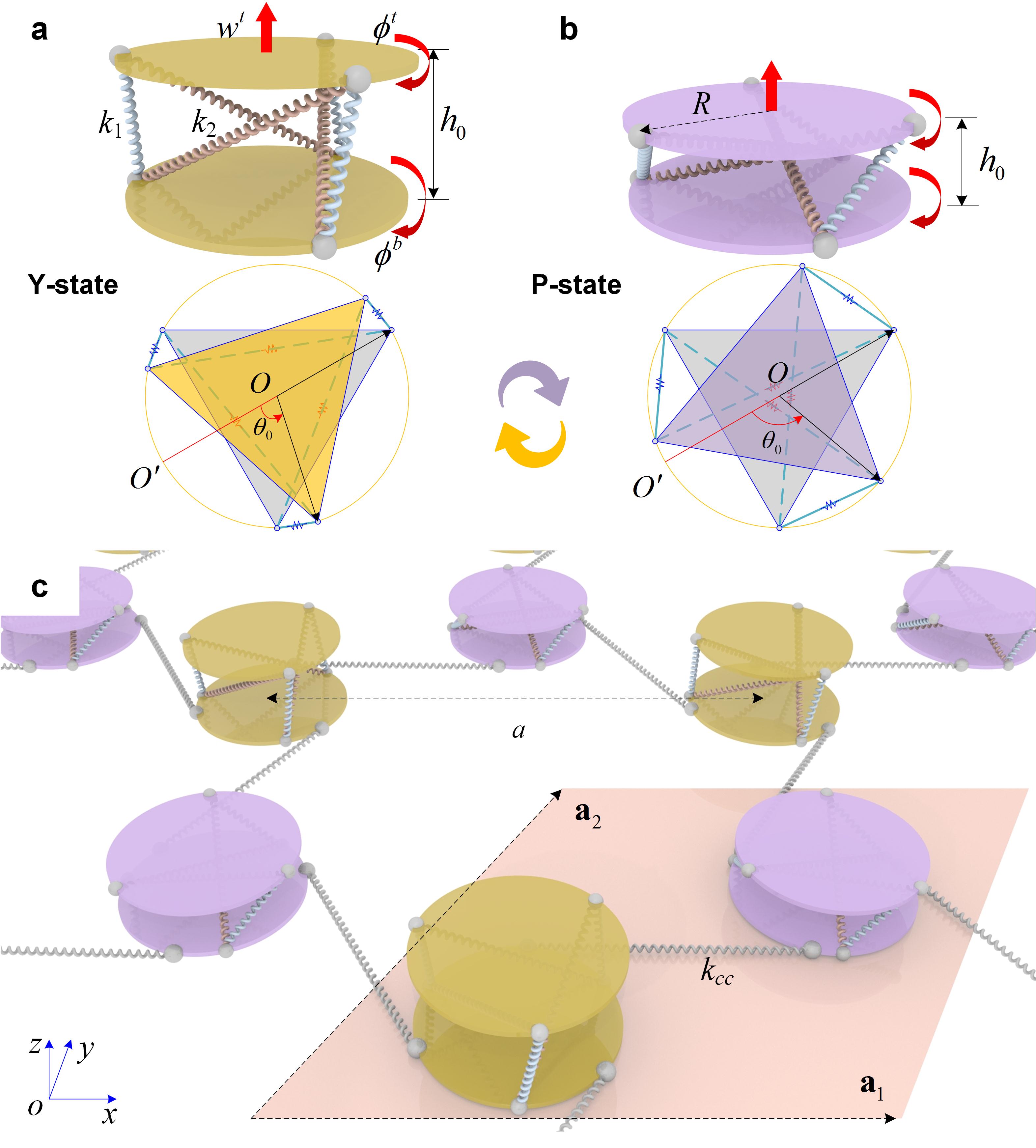}
\caption{\textbf{Design of the tunable topological metamaterial based on bistable SP.} (\textbf{a})-(\textbf{b}) The two stable states, Y- and P-states, and below are their top views.  They differ in the height ($h_0$) and the relative rotation ($\theta_0$) between the disks: Y-state ($h_0 = 0.04$ m, $\theta_0= 70^{\circ}$) and P-state ($h_0 = 0.0215$ m, $\theta_0 = 110^{\circ}$). Disk radius $R = 0.0408$ m, connecting spring coefficients $k_1 = 10^2$ N/m and $k_2 = 10^5$ N/m. Three allowable degrees of freedom, $\phi^b$, $\phi^t$, and $w^t$, are marked with the red arrows. (\textbf{c}) Hexagonal arrangement by combining the two stable states. The lattice constant $a$ of the hexagonal arrangement is 0.212 m, and the coefficient $k_{cc}$ of the reverse spring is $10^2$ N$\cdot$m. The periodic unit cell consisting of one Y-state and one P-state is highlighted along with its basis vectors $\mathbf{a}_1$ and $\mathbf{a}_2$.
}
\label{fig1}
\end{figure} 

\textbf{Design of the tunable topological metamaterial.} The tunable system we propose is illustrated in Fig.~\ref{fig1}. Each SP is bistable, i.e., it has two stable configurations: Y-state (yellow) and P-state (purple) as shown in Fig.~\ref{fig1}a-b. This unit is made of two parallel disks connected with six linear springs. The conventional SP unit has six DOFs for the top disk and the bottom disk is fixed \citep{41}. However, by judiciously choosing the connecting springs, one can decouple some DOFs and reduce the total DOFs (see Supplementary Note 1). In this study, for the sake of simplicity, we assume that the bottom disk in pinned at its center, such that it can only rotate about the $z$-direction. We denote this rotational DOF of the bottom disk by $\phi^{b}$. The top disk can have only rotational ($\phi^t$) and translational ($w^t$) motions along the $z$-direction with respect to its equilibrium position. Note that these dynamic perturbation parameters, $\phi^{b}$, $\phi^t$, and $w^t$, should not be confused with $\theta_0$ and $h_0$, which denote the equilibrium parameters of the SP unit cell and vary depending on whether it is in the Y- or P-state (Fig.~\ref{fig1}a). All three DOFs of these dynamic motions in terms of $\phi^{b}$, $\phi^{t}$, and $w^{t}$ are governed by the six springs  between the plates. The detailed mathematical relationships -- including the derivations of the bistability in the SP unit cell -- are described in Supplementary Figure 1 and Note 2.


We design a hexagonal lattice by combining two stable states, such that the system breaks $C_6$ symmetry but retains $C_3$ symmetry (Fig.~\ref{fig1}c). Only the bottom disk of each SP is connected with neighbouring SPs with springs. Note that this connection is a reverse spring, i.e., it induces opposite torque in the connected units (a similar setup can be found in~\citep{34}). Tunability comes from the fact that one can easily change the stable state of each SP---independently---to achieve a desired lattice configuration.

\vspace{3ex}
\textbf{Band-inversion and topology.} In this section, we evaluate dispersion characteristics of the system and observe a topological transition. For describing the dynamics of the (infinite) hexagonal lattice, we  choose a periodic unit cell (highlighted in Fig.~\ref{fig1}c), which consists of two SPs indexed as 1 and 2. Here, each SP can take either Y- or P-state by dial-in actions. Since a single SP has three DOFs, the unit cell would be represented by the following six parameters: [$\phi^{(1)b}$, $\phi^{(2)b}$, $\phi^{(1)t}$, $\phi^{(2)t}$, $w^{(1)t}$, $w^{(2)t}$]. 

\begin{figure}[t]
\centering
\includegraphics[width=6in]{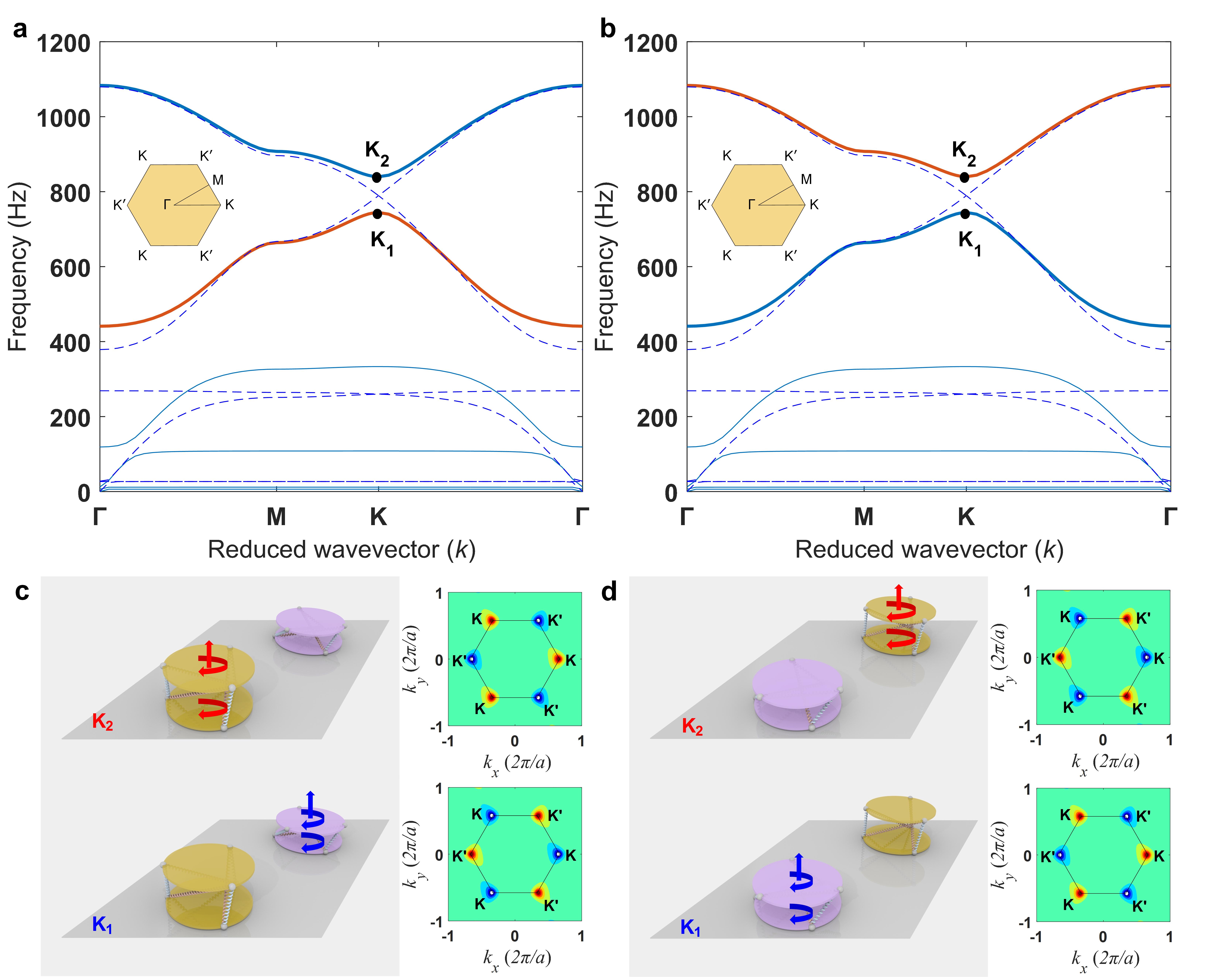}
\caption{\textbf{Band-inversion when the unit cell is transformed from the Y-P to the P-Y configuration.} (\textbf{a}) Unit cell dispersion relation for the Y-P configuration. A complete band gap emerges around 800 Hz due to broken inversion symmetry. The dashed lines with a Dirac cone at the \textbf{K} point represent the case with both SPs of the unit cell having the same effective stiffness. (\textbf{b}) The same for the P-Y state.  Left panels in (\textbf{c}) and (\textbf{d}) represent the eigenmodes corresponding to \textbf{K$_1$} and \textbf{K$_2$} points in the dispersion relations of the Y-P and the P-Y configurations, respectively. Only one SP shows the motion, demonstrating a spatial inversion in the unit cell for the two configurations. In the right panels of (\textbf{c}) and (\textbf{d}), this reflects as the change in the sign of the Berry curvature at the \textbf{K} and \textbf{K$'$} points in the first Brillouin zone (solid line). }
\label{fig2}
\end{figure}

First, we write the equations of motion for the periodic unit cell (index as $i, j$) as: 
\begin{subequations}
\begin{eqnarray}
  I\ddot \phi _{i,j}^{(1)b} + {k_{cc}}(3\phi _{i,j}^{(1)b} + \phi _{i,j}^{(2)b} + \phi _{i,j - 1}^{(2)b} + \phi _{i - 1,j}^{(2)b}) - k_{\phi w}^{(1)}w_{i,j}^{(1)t} + k_{\phi \phi }^{(1)}(\phi _{i,j}^{(1)b} - \phi _{i,j}^{(1)t}) = 0 \\ 
  I\ddot \phi _{i,j}^{(1)t} + k_{\phi w}^{(1)}w_{i,j}^{(1)t} + k_{\phi \phi }^{(1)}(\phi _{i,j}^{(1)t} - \phi _{i,j}^{(1)b}) = 0 \\ 
  m\ddot w_{i,j}^{(1)t} + k_{ww}^{(1)}w_{i,j}^{(1)t} + k_{w\phi }^{(1)}(\phi _{i,j}^{(1)t} - \phi _{i,j}^{(1)b}) = 0 
 \end{eqnarray}
 \end{subequations}
 \begin{subequations}
\begin{eqnarray}
  I\ddot \phi _{i,j}^{(2)b} + {k_{cc}}(3\phi _{i,j}^{(2)b} + \phi _{i,j}^{(1)b} + \phi _{i,j + 1}^{(1)b} + \phi _{i + 1,j}^{(1)b}) - k_{\phi w}^{(2)}w_{i,j}^{(2)t} + k_{\phi \phi }^{(2)}(\phi _{i,j}^{(2)b} - \phi _{i,j}^{(2)t}) = 0 \\ 
  I\ddot \phi _{i,j}^{(2)t} + k_{\phi w}^{(2)}w_{i,j}^{(2)t} + k_{\phi \phi }^{(2)}(\phi _{i,j}^{(2)t} - \phi _{i,j}^{(2)b}) = 0 \\ 
  m\ddot w_{i,j}^{(2)t} + k_{ww}^{(2)}w_{i,j}^{(2)t} + k_{w\phi }^{(2)}(\phi _{i,j}^{(2)t} - \phi _{i,j}^{(2)b}) = 0 
\end{eqnarray}
 \end{subequations}


\noindent where $I$ and $m$ are the rotational inertia and the mass of the disks, respectively. $k_{ww}$, $k_{\phi \phi}$, $k_{w \phi}$, and $k_{\phi w}$ are the stiffness coefficients for the relative translation and rotations, and their coupling. The detailed expressions of these coefficients as a function of $k_1$, $k_2$, and geometric parameters are described in Supplementary Note 1.

For a lattice length of $a$, we invoke the Bloch's theorem by using the periodicity in two directions: $\mathbf{a}_1=[1,0]a$ and $\mathbf{a}_2=[1/2,\sqrt{3}/2]a$, and obtain the following eigenvalue problem:
\begin{eqnarray}
\omega^2 \mathbf{M} \mathbf{U}= \mathbf{B}\mathbf{U}
\end{eqnarray}
where $\omega$ is the angular frequency, and the generalized eigenvector $\mathbf{U}=[\phi_{i,j}^{(1)b}, \phi_{i,j}^{(2)b}, \phi_{i,j}^{(1)t}, \phi_{i,j}^{(2)t}, w_{i,j}^{(1)t}, w_{i,j}^{(2)t}]$. $\mathbf{B}$ and $\mathbf{M}$ are the stiffness and mass matrices, respectively (see the detailed expressions in Supplementary Note 3).

There are two possible configurations of the $C_3$ symmetric hexagonal unit cell: Y-P configuration and P-Y configuration. Figure~\ref{fig2} displays the dispersion properties and associated band-inversion when one state is transformed to the other. Due to the breakage of the inversion symmetry, a complete band gap emerges at the \textbf{K} point between the fifth and the sixth bands (see the highlighted red and blue bands in Fig.~\ref{fig2}a-b). Note that these bands would have a degeneracy at the \textbf{K} point if the inversion symmetry is not broken (see the blue dashed lines). Although the dispersion curves for these two Y-P and P-Y configurations look similar, there is a difference in terms of the topology. The highlighted bands are inverted in these configurations --- so-called band-inversion. We verify this by plotting in the left columns of Figs.~\ref{fig2}c-d the mode shapes of the unit cell corresponding to the points \textbf{K$_1$} (744.5 Hz) and \textbf{K$_2$} (839.8 Hz) marked in the dispersion relations. For the Y-P configuration in Fig.~\ref{fig2}c, we see that the low frequency vibration (\textbf{K$_1$} point) corresponds to the case when only the P-state is vibrating in the lattice. In Fig.~\ref{fig2}d, the P-state vibrates in the P-Y configuration at the low frequency in the same way. This makes sense as the designed stiffness for the P-state is lower than that of the Y-state (see Supplementary Note 2). However, there is an inversion in terms of where in the unit cell the vibration is dominant due to the filpping of the Y- and P-states.  

The aforementioned band-inversion resembles the one seen in the valley Hall effect. We can calculate the valley Chern number in order to track the topological transition associated with this effect. This is achieved by integrating its Berry curvature over \textit{half} of the first Brillouin zone. Mathematically, the valley Chern number $C=(1/2\pi) \iint \Omega(\mathbf{k})dk $, where the Berry curvature $\Omega(\mathbf{k})=i \sum_{v=1, v \neq u }^{6} \left[ \langle \mathbf{u} | \partial{\mathbf{B}}/\partial{k}_x| \mathbf{v} \rangle  \langle \mathbf{v} | \partial{\mathbf{B}}/\partial{k}_y| \mathbf{u} \rangle - c.c. \right]/ (\omega_v^2-\omega_u^2)^2$ \citep{39}. Here, \textit{i} is the imaginary unit, $c.c.$ denotes complex conjugate, $\langle \cdot | \cdot \rangle$ represents the inner product, $\mathbf{k}=[k_x, k_y]$ is the wave vector, and $\omega_u$ and $\omega_v$ are the angular frequencies corresponding to the normalized eigen vectors $\mathbf{u}$ and $\mathbf{v}$, respectively. The calculated Berry curvature in the first Brillouin zone is shown in the right columns of Figs.~\ref{fig2}c-d. One notices that it is localized at \textbf{K} and \textbf{K$'$} points in the Brillouin zone, and it changes its sign as we alter from the Y-P to the P-Y configuration---reflecting the band-inversion process. Therefore, the calculated valley Chern numbers for the fifth and the sixth bands of the Y-P configuration (highlighted red and blue bands in Fig.~\ref{fig2}a) are $-1/2$ and 1/2, respectively. These are reversed for the P-Y configuration, and this confirms that the two configurations are topologically distinct. The quantized difference of the valley Chern numbers of the two configurations, i.e., $|C_{\text{Y-P}}-C_{\text{P-Y}}|=1$, indicates the emergence of a topologically protected edge state at the interface if these configurations are placed adjacently. 
  
\vspace{3ex}

\begin{figure}[t]
\centering
\includegraphics[width=6in]{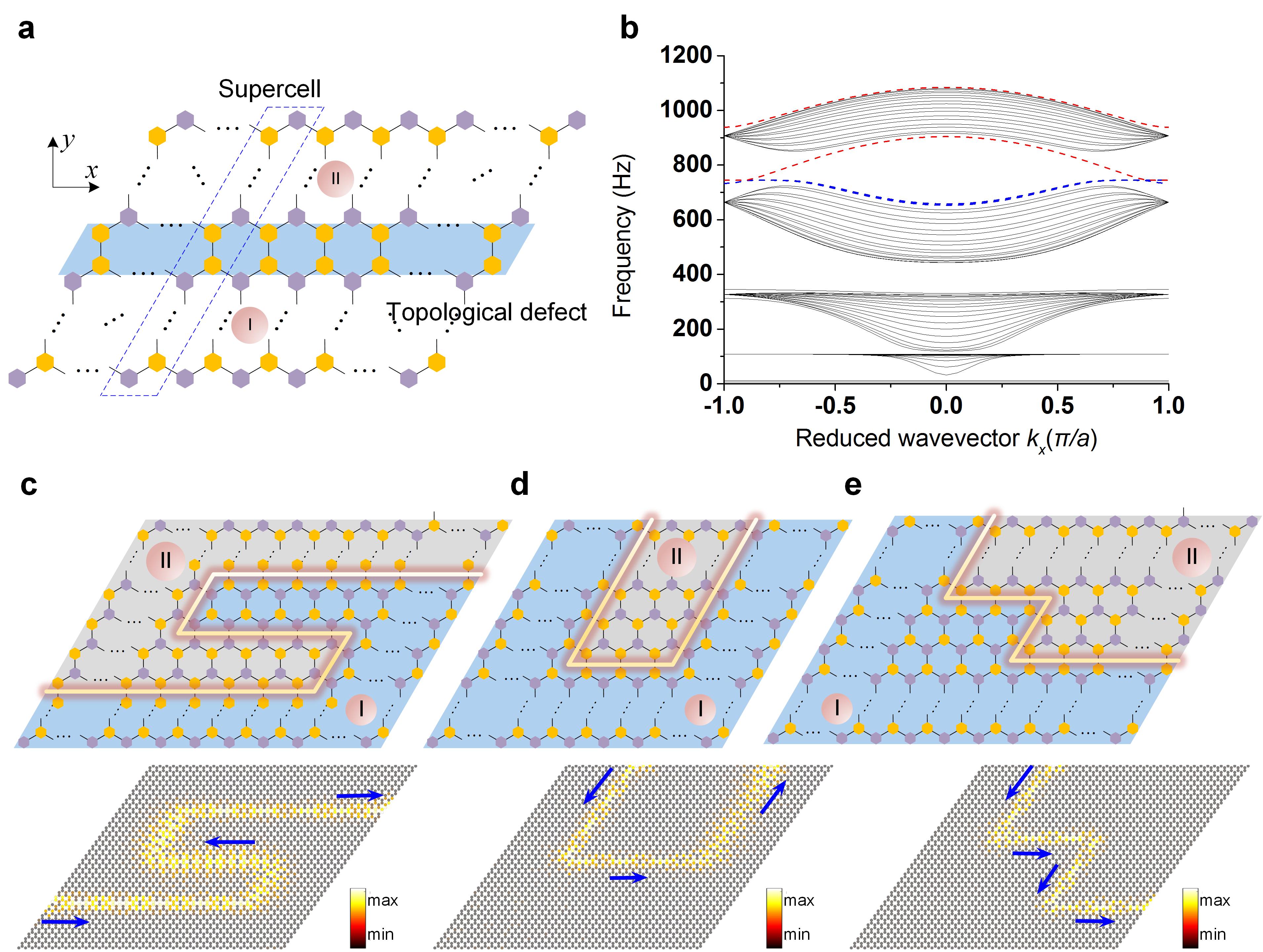}
\caption{\textbf{Topologically protected interface mode between two topologically distinct lattices, and waveguide path manipulation via dial-in actions.} (\textbf{a}) A linear topological defect is highlighted by the blue background at the interface of the P-Y (I) and the Y-P (II) configurations. Dynamic properties are evaluated by taking a supercell enclosed by the dashed box. (\textbf{b}) Band structure for the supercell. The dashed red and blue curves represent the localized modes in the supercell. (\textbf{c})-(\textbf{e}) A wide range of waveguide paths at the intersection of I and II can be achieved simply by dial-in actions in the lattice. Below are the numerical simulations to confirm the robust energy transport along these interfaces. Colors denote the rotation of the bottom disks.}
\label{fig3}
\end{figure}

\textbf{Topological defect and its manipulation.} In this section, we will show how a topological defect can be created by placing topologically distinct lattices, P-Y (defined hereafter to be type-I for the sake of simplicity) and Y-P (type-II), adjacently. We will also show that this topological defect can  easily be reconfigured to any other shapes---thanks to the extreme tunablity of the system. First, we confirm the existence of topologically protected interface modes for a linear topological defect (see Fig.~\ref{fig3}a). To this end, we take a supercell of size $1 \times 20$ with the top and the bottom boundaries fixed, and we apply the periodic boundary condition in the $x$-direction. 

The resulting dispersion of the supercell strip is plotted in Fig.~\ref{fig3}b. We notice several bulk modes (curves in black color). These correspond to the bulk bands seen in the unit cell analysis done earlier (Figs.~\ref{fig2}a-b). However, we also note some additional modes (dashed red and blue curves in Fig.~\ref{fig3}b). In blue color are the two overlapping modes that appear at the top and the bottom of the supercell due to the identical boundaries on both sides. However, there is a distinguished mode (red color) \textit{inside} the band gap. This corresponds to the interface mode emerging due to the distinct topological nature of the lattices I and II. Additionally, to the extreme top of the dispersion curve, we do observe another interface mode (red color) outside the cutoff frequency. 

Now that we have shown the framework of creating a topological defect, it is also possible to achieve various complex shapes of the topological interfaces in a 2D lattice by strategic dial-in actions of the SP cells. In Figs.~\ref{fig3}c-e, we assemble some of these shapes with various bends, showcasing the manipulation capability in the system. Below are the plots, obtained from the numerical experiments (see methods) for a harmonic excitation at 760 Hz applied to the 40 $\times$ 40 lattice. This frequency excitation, being inside the band gap, excites the topologically protected mode, and thus demonstrates a robust energy transport through various interfaces.

\vspace{3ex}

\begin{figure}[!]
\centering
\includegraphics[width=6in]{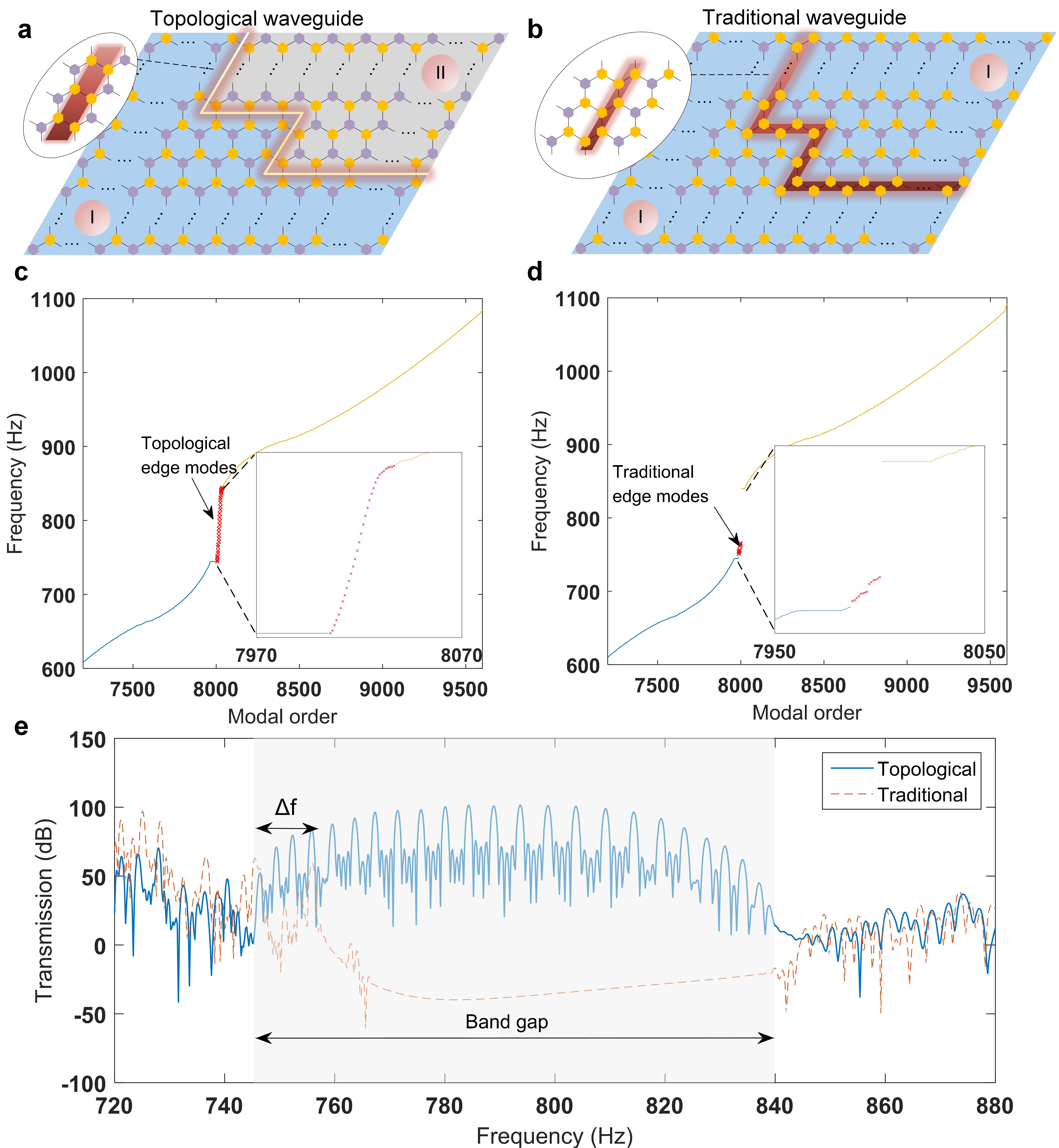}
\caption{\textbf{Comparison between topological and traditional waveguides.} (\textbf{a})-(\textbf{b}) A finite $40 \times 40$ lattice can either be dialed in to form a topological or a traditional waveguide. (\textbf{c})-(\textbf{d}) Eigenfrequencies of these lattices arranged in the modal order. In (\textbf{c}), the entire band gap is populated with topologically protected edge modes at the interface (zoomed-in view in the lower inset). In (\textbf{d}), the traditional waveguide does not lead to the edge modes spanned over the entire band gap (zoomed-in view in the lower inset). (\textbf{e}) Transmission spectrum of both waveguides.}
\label{fig4}
\end{figure}

\textbf{Topological vs. traditional waveguides.} It is now natural to ask --- in what ways the \textit{topological} waveguide different from a \textit{traditional} waveguide? Here, by a traditional waveguide, we refer to a waveguide that is created without incurring topological disparities in the same system. 
One way to create such a waveguide is by introducing a \textit{topologically trivial} defect along the desired wave path so as to utilize the localized defect modes for wave transmission \citep{43}. Figures~\ref{fig4}a-b show the exemplary cases of topological and traditional waveguides, respectively, realized in the same SP settings. Note that given the initially uniform $C_3$ symmetric hexagonal structure, we can achieve the creation of the traditional defect simply by transforming P-states into Y-states (i.e., \textit{in situ} dial-in action) along the desired wave path (see the inset of Fig.~\ref{fig4}b). However, the creation of the topological defects in the originally uniform lattice requires more actions, since it necessitates the border between the type-I and the type-II lattices. That is, we need to convert the one side of the SP cells (i.e., the upper side of the lattice with respect to the wave path in Fig.~\ref{fig4}a) from the type-I to the type-II lattice, which demands the whole flipping (i.e., dial-in action) of the SP cells from the Y- to P-state and vice versa. The analogous operation in 1D lattice systems has been investigated by Chaunsali \textit{et al}. \citep{38}. The advantage of our system is that it can realize both types of traditional and topological waveguides in the same system by leveraging the bistable SP network and the \textit{in situ} dial-in action on it. This provides an excellent opportunity to compare their transmission properties. 

We perform an eigen analysis on a $40\times40$ structure (with 9600 DOF in total) for both types of waveguide structures. In Figs.~\ref{fig4}c-d, we plot the eigen frequencies against the modal order for the topological and traditional waveguides, respectively.  The blue and yellow curves correspond to the bulk bands, highlighted in the unit cell dispersion in Figs.~\ref{fig2}a-b. Sandwiched between these two branches are the modes localized at the waveguide interfaces (red color, as magnified in the insets). We observe that the topological waveguide ensures that the entire band gap is populated by the interface modes. This affirms our observations in the previous sections where a robust transmission was shown along topological interfaces. The traditional waveguide, however, does not lead to the modes spanned over the entire band gap (Fig.~\ref{fig4}d)---indicating its limitations in terms of constructing a wide range of defect modes. For the corresponding eigen shapes, see Supplementary Movies 1 and 2.

The aforementioned characteristics provide us with a deep insight into the differences of the topological and traditional waveguides, and these are now used to explain different transmission spectra along the waveguide channels. We perform numerical simulations and calculate the transmission spectra for a range of input frequencies (see methods). In Fig.~\ref{fig4}e, we plot the obtained transmission. We can immediately notice a clear difference in the transmission inside the band gap (i.e., 744.5 Hz to 839.8 Hz). The topological waveguide leads to a superior transmission all along the band gap, however, the traditional waveguide yields a significant transmission only in a small range of frequencies ($\Delta f$). This transmission limited to a small frequency window is due to the presence of the defect modes, previously shown in Fig.~\ref{fig4}d. 
\begin{figure}[!]
\centering
\includegraphics[width=6in]{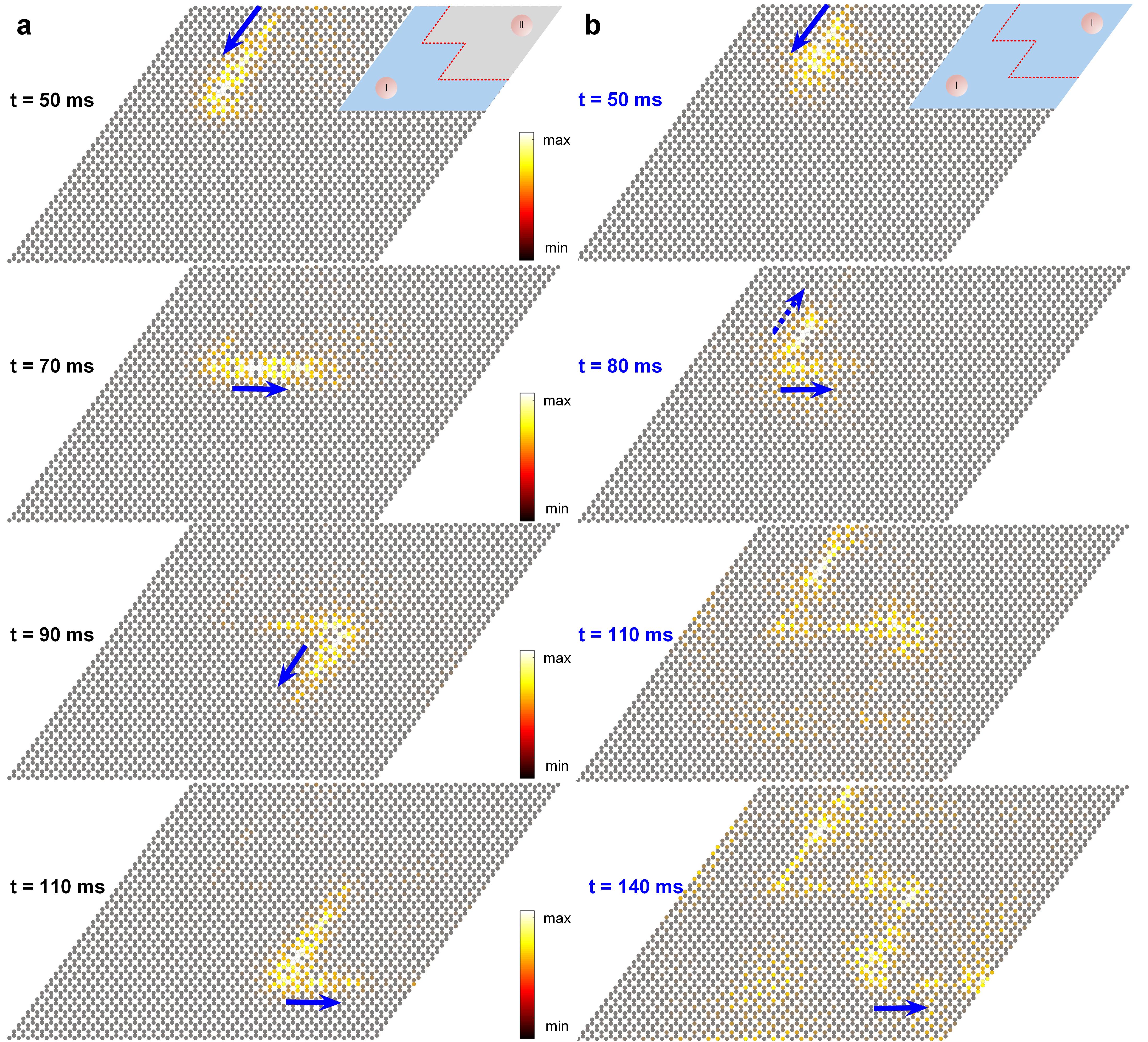}
\caption{\textbf{Wave packet scattering} (\textbf{a}) A topological waveguide showing back scattering suppressed wave propagation. (\textbf{b}) The same for a traditional waveguide. Though the wave packet reaches the exit, its propagation is not scattering free. Colors denote the rotation of the bottom disks.}
\label{fig5}
\end{figure} 

Even though the topological waveguide shows a superior transmission overall for the frequencies inside the band gap, we investigate if it has any \textit{qualitative} differences from the traditional waveguide for the small range of frequencies ($\Delta f$ in Fig.~\ref{fig4}e). To answer this question, we perform a numerical experiment on the waveguides by sending a 50 ms long Gaussian packet at a frequency (757 Hz) lying inside the range of interest $\Delta f$. In Fig.~\ref{fig5}, we compare the transient responses for topological and traditional waveguides. We observe that the topological waveguide in Fig.~\ref{fig5}a shows that the wave packet does not back scatter at the multiple sharp bends, and it smoothly travels along the path. However, the traditional waveguide in Fig.~\ref{fig5}b indicates the presence of scattering around the bends. Note that the wave packet still propagates along the path and would lead to a significant transmission as shown in Fig.~\ref{fig4}e. Nonetheless, it is qualitatively different from the topological waveguide in terms of guiding a wave packet by allowing back scattering around bends. The difference of wave transmission efficiency between the traditional and topological waveguides is further investigated systematically for bends of various angles in Supplementary Figures 2 $\sim$ 4 and Note 4. 
 
\vspace{3ex}

\begin{figure}[!]
\centering
\includegraphics[width=6in]{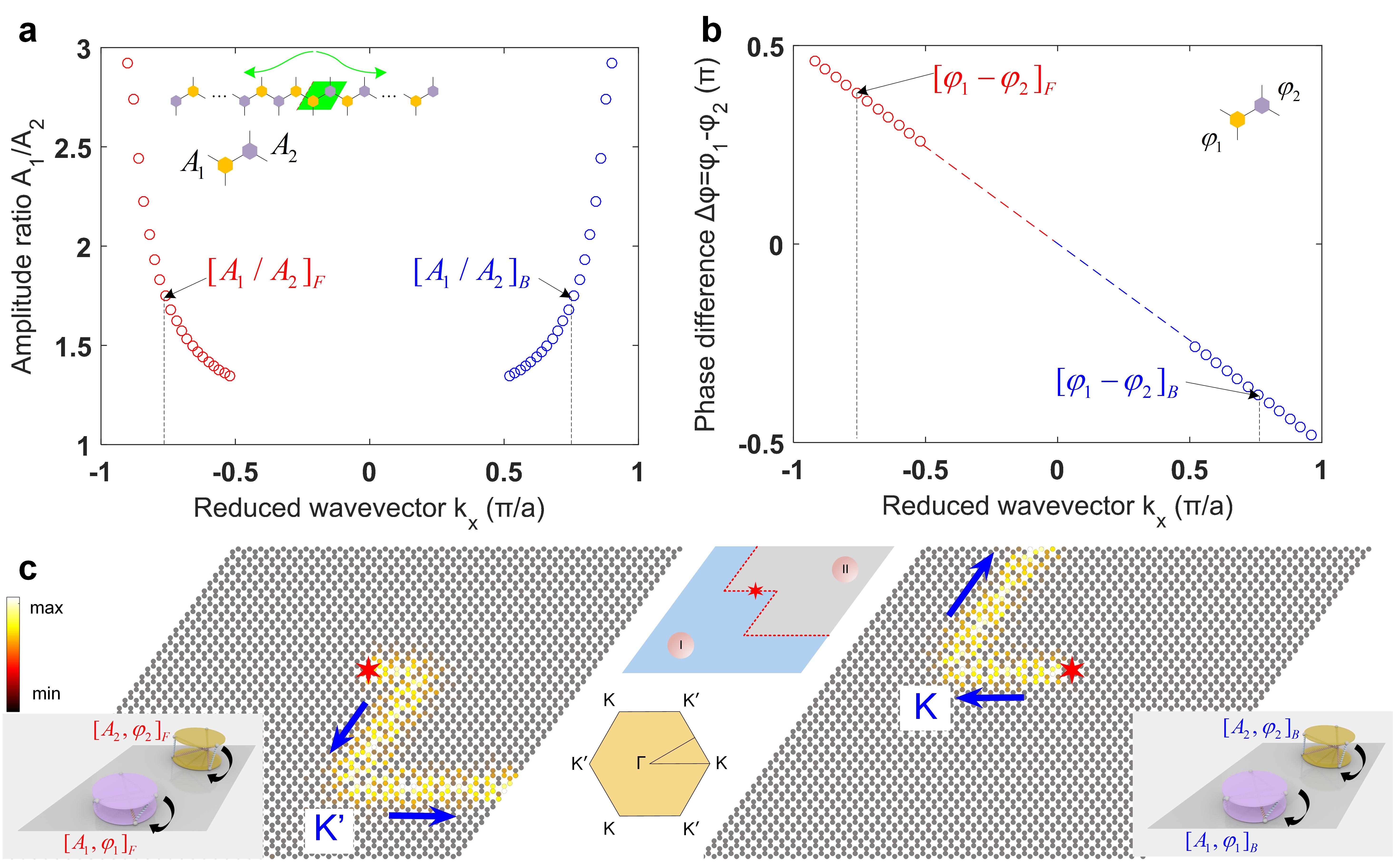}
\caption{\textbf{Valley-selective one-way wave propagation} (\textbf{a}) The extracted amplitude ratio ($A_1/A_2$) of Y- and P-states in the unit cell (green box in inset) from the supercell analysis. The data points on the left side (red dots) correspond to the solution with positive group velocity, i.e, forward propagation, while the right ones (blue dots) are for backward propagating wave. These are the solutions for two valleys. (\textbf{b}) Phase difference (${\varphi _1} - {\varphi _2}$) extracted for the same states. (\textbf{c}) Numerical simulation results that show one-way wave propagation along the topological waveguide. The red star indicates the place where the valley-selective excitation is given (see insets) to induce either forward-  or backward-propagating waves. Colors denote the rotation of the bottom disks.}
\label{fig6}
\end{figure}

\textbf{One-way waveguide.} Lastly, we demonstrate that a topological waveguide constructed through our system can support one-way wave propagation if a valley-selective excitation is given at the source. To this end, we first extract the amplitude and phase information of the topologically protected mode from the supercell analysis done earlier (Figs.~\ref{fig3}a-b). In Figs.~\ref{fig6}a-b, we calculate the amplitude ratio (i.e., $A_1/A_2$) and phase difference (${\varphi _1} - {\varphi _2}$) of the bottom disks' rotations between the Y-and P-states at the topological interface (see insets), and plot them as a function of reduced wavevector ($k_x$) in the bandgap range. We observe that the trend can be categorized into two \textit{valleys}. One corresponds to the forward propagation (data points indicated with the subscript \textit{F} in Figs.~\ref{fig6}a-b), and the other to the backward propagation (subscript \textit{B}). These can be verified by looking at the slope (i.e., group velocity) of the protected mode (red color) in Fig.~\ref{fig3}b. In this way, appropriate amplitude ratio and phase difference can be applied (i.e., valley-selection at either $\mathbf{K}$ or $\mathbf{K'}$ ) to the disks at the interface to excite either forward or backward propagating waves. This is confirmed by a transient analysis performed at 778 Hz in Fig.~\ref{fig6}c. If the lattice is excited in the middle point (marked with a star) with the amplitude of $A_1/A_2 = 1.748$ and the phase difference of ${\varphi _1} - {\varphi _2} = 0.38\pi$, we observe a forward propagating wave (left panel in Fig.~\ref{fig6}c). However, if the excitation with the same amplitude but opposite phase (i.e., ${\varphi _1} - {\varphi _2} = -0.38\pi$) is applied, we observe a backward propagating wave (right panel). See Supplementary Movies 3 and 4 for more details. These numerical results attest that the bistable SP-based metamaterial system proposed in this study  allows not only the \textit{in situ} manipulation of wave paths via dial-in actions, but also a selective one-way propagation of mechanical waves by strategic excitations. 
 
\section*{DISCUSSION}
Here, we propose a dial-in topological metamaterial system based on the bistable Stewart Platform (SP) and report robust one-way propagation of mechanical waves in it along tailorable wave paths. By arranging the bistable SP cells hexagonally in an alternating fashion, we can create two types of topologically distinctive lattices, which can be transformed to each other simply by a dial-in action. We prove that this transformation changes the topology of the system---quantified by the valley Chern numbers. When lattices with these two topologically distinct configurations are placed adjacently, we show the existence of a topologically protected mode at the interface. This idea is extended to tune the system \textit{in situ} to create variety of waveguides, and we demonstrate a robust energy propagation along them using numerical simulations. We conduct eigenmode analysis and transient simulations of finite structures to highlight some key differences between topological and traditional waveguides in the same system. While the traditional waveguides also lead to significant wave transmission due to interfacial local modes inside the band gap, the edge modes generated in the topological waveguides are qualitatively different in that they support a wider range of frequencies and are immune to back scattering at sharp bends in the structure. We also show the strategy of giving valley-selective excitation to the system such that one-way wave propagation is achieved along the waveguide. Therefore, this tunable system opens up possibilities to realize various complex shapes of topological waveguides without resorting to external fields or adding/removing the masses from the system. Further studies including the experimental verification of the proposed tunable metamaterials would be reported in authors' future publications. 

\section*{METHODS}

\textbf{Numerical experiment:} We employ the Runge-Kutta method (step size $=10^{-4}$ s) to get the response at any time instant for a variety of input signals: harmonic, Gaussian pulse, and sinusoidal frequency sweep.

\textbf{Calculation of the transmission spectrum:} We perform a numerical experiment and calculate the transmission spectra. A sweep frequency signal ($20$ Hz to $1000$ Hz in 5 s) is used as a rotational perturbation applied to the bottom disk at the input location. The transient rotation of the bottom disk is measured at the output location. Thus, the transmission $T(\omega)=20 \log[\Phi_{output}^{b}(\omega)/\Phi_{input}^{b}(\omega)]$, where $\Phi(\omega)$ represents the power spectral density (PSD) of the transient rotation $\phi(t)$ of the disk.
 
\section*{DATA AVAILABILITY}
The data that support the findings of this study are available from the corresponding author upon request.

\section*{ACKNOWLEDGEMENTS}
The authors are grateful to Rui Zhu, Linyun Yang, and Xiaotian Shi for fruitful discussions. R. C., H. Y., and J. Y. are grateful for the financial support from the U.S. National Science Foundation (CAREER-1553202 and EFRI-1741685). Y. W. and K. Y acknowledge the support from the Harbin Institute of Technology and the China Scholarship Council (Grant No. 201606120065). 

\section*{AUTHOR CONTRIBUTIONS}
Y. W., J. Y., and K. Y. conceived the original idea. Y. W. performed the numerical simulations. H. Y. calculated the bistable characteristic of SP. Y. W. and R. C. analyzed the results and wrote the manuscript. J. Y. and K. Y. supervised the project.

\section*{ADDITIONAL INFORMATION}
\textbf{Competing financial interests:} The authors declare no competing financial interests.

\end{document}